\documentclass[a4paper,11pt]{article}
\pdfoutput=1 

\usepackage{jheppub} 

\usepackage[T1]{fontenc} 
\usepackage{graphicx}
\usepackage{subcaption}
\usepackage{slashed}
\usepackage{braket}
\usepackage{bbold}
\usepackage{orcidlink}
\usepackage{bbding}
\usepackage[normalem]{ulem}

\usepackage[compress]{natbib}
\setcitestyle{square,numbers,comma}

\title{\boldmath Constraints on NJL four-fermion effective interactions from neutrinoless double beta decay}

\author[a,b]{L. Pacioselli\orcidlink{0009-0000-3942-0721},}
\author[b,1]{O. Panella\orcidlink{0000-0003-4262-894X},\note{Corresponding author.}}
\author[c]{M. Presilla\orcidlink{0000-0003-2808-7315}}
\author[b,d,e,f]{and S.-S. Xue\orcidlink{0000-0002-2702-0859}}

\affiliation[a]{Dipartimento di Fisica e Geologia, Universit\`{a} degli Studi di Perugia, Via A. Pascoli, I-06123, Perugia, Italy}
\affiliation[b]{INFN Sezione di Perugia, Via A. Pascoli, I-06123, Perugia, Italy}
\affiliation[c]{Institute for Experimental Particle Physics (ETP), Karlsruhe Institute of Technology (KIT), Wolfgang-Gaede-Straße 1, 76131 Karlsruhe, Germany}
\affiliation[d]{ICRANet, Piazzale della Repubblica, 10-65122, Pescara, Italy}
\affiliation[e]{Dipartimento di Fisica, Sapienza-Università di Roma, Piazzale Aldo Moro 5, 00185 Roma, Italy}
\affiliation[f]{ICTP-AP, University of Chinese Academy of Sciences, Beijing, China}

\emailAdd{luca.pacioselli@pg.infn.it}
\emailAdd{orlando.panella@cern.ch}
\emailAdd{matteo.presilla@cern.ch}
\emailAdd{xue@icra.it}

\abstract{We study the contribution of a heavy right-handed Majorana neutrino to neutrinoless double beta decay ($0\nu\beta\beta$) via four-fermion effective interactions of Nambu-Jona-Lasinio (NJL) type. In this physical scenario, the sterile neutrino contributes to the nuclear transition through gauge, contact, and mixed interactions. Using the lower limit on the half-life of $0\nu\beta\beta$ from the KamLAND-Zen experiment, we then constrain the effective right-handed coupling between the sterile neutrino and the $W$ boson: $\mathcal{G}^{W}_{R}$. Eventually, we show that the obtained bounds are compatible with those found in the literature, which highlights the complementarity of this type of phenomenological study with high-energy experiments.}

\begin{document} 
\maketitle
\flushbottom
\section{Introduction}
\label{sec:intro}

The Standard Model (SM) for elementary particle physics is the experimentally confirmed~\cite{Wu:1957my,CMS:2012qbp,ATLAS:2012yve} gauge theory that led to the most accurate results in Physics history. Nevertheless, SM alone cannot answer many questions raised by experimental evidence: the existence of dark matter, General Relativity, the baryon asymmetry problem, the hierarchy pattern of fermion masses, etc. For these reasons, a multitude of Beyond the Standard Model (BSM) theories have been developed and tested in various experimental facilities, e.g. the searches for contact interactions at the Large Hadron Collider (LHC)~\cite{ATLAS:2017eqx,CMS:2018ucw} and the searches for lepton number violation (LNV) at low-energy experiments~\cite{GERDA:2020xhi,CUORE:2017tlq,KamLAND-Zen:2022tow}.

The most interesting physical process for detecting LNV signals is the neutrinoless double beta decay ($0\nu\beta\beta$)~\cite{Furry:1939qr,Zeldovich:1981da,Agostini:2022zub}: a rare nuclear decay strictly forbidden by the SM that, due to the Schechter-Valle theorem~\cite{Schechter:1981bd}, would imply the Majorana nature for neutrinos. The first direct searches for $0\nu\beta\beta$ decays began in the 1960s~\cite{derMateosian:1964vza,Fiorini:1967in}, but it has never been observed to date.  However, the expected sensitivity gain of the next-generation experiments looks promising~\cite{LEGEND:2021bnm,nEXO:2021ujk,McDonald:2017izm,Nakamura:2020szx}. Hence, from these experiments are derived only lower limits for the half-life $T^{0\nu}_{1/2}$. The aim of this paper is to exploit those experimental limits in order to constrain the parameters of a specific class of BSM models, as extensively done in the literature~\cite{Blennow:2010th,Mitra:2011qr,Asaka:2013jfa,Barea:2009zza,Panella:1997wa,Biondini:2021vip,Faessler:2014kka}. 

At low energies, the SM with one elementary Higgs boson~\cite{Higgs:1964pj} is rendered, as an effective field theory (EFT), by the Nambu-Jona-Lasinio (NJL) model~\cite{Nambu:1961fr,Nambu:1961tp} of four-fermion interactions, where the Higgs is a composite particle. 
In this paper, we consider the latter, in which the effective Lagrangian contains four-fermion operators responsible, through minimal dynamical symmetry breaking in a well-defined quantum field theory (QFT) at the high energy scale, for generating the top quark and Higgs masses~\cite{Bardeen:1989ds, Xue2013}. This model has been used to tackle several open questions of the SM: from the hierarchy pattern of fermion masses~\cite{Xue:2016dpl} to dark matter particles' masses~\cite{Xue:2020cnw}, and the recent $W$ boson mass tension~\cite{Xue2022b}. The model not only renders the effective parity-violating SM at its infrared (IR) fixed point of the electroweak scale $v = 246 \text{ GeV}$, but also a parity-preserving theory of massive composite particles at its ultraviolet (UV) fixed point of the composite scale $\Lambda\sim \mathcal{O}$(TeV) scales \cite{Xue2017}. Hence, the $W$ boson gauge coupling is no longer purely left but has a non-trivial right coupling from the four-fermion interactions at high energies. We will study and constrain precisely this effective right coupling $\mathcal{G}^W_R$ through this $0\nu\beta\beta$-driven study, i.e. at low energies ($\sim 100 \text{ MeV}$).

The rest of the paper is organized as follows: in Sec.~\ref{sec:NJL} we outline the main features of the four-fermion interactions of the NJL-type model employed; in Sec.~\ref{sec:computations} we use the model interactions contributing to $0\nu\beta\beta$ (gauge, contact and two mixed) to compute its theoretical half-life; then in Sec.~\ref{sec:results} we exploit the experimental lower limits on $T^{0\nu}_{1/2}$ to extract the bounds on the new composite scale $\Lambda$ and effective right-handed coupling $\mathcal{G}^W_R$ of the model; lastly, we summarize the work with some remarks in Sec.~\ref{sec:conclusion}.

\section{Four-fermion interactions of NJL type}
\label{sec:NJL}

Here we briefly describe the model adopted in this work. As shown in low-energy experiments, the SM possesses parity-violating (chiral) gauge symmetries $SU_c(3)\times SU_L(2)\times U_Y(1)$. 
As a well-defined QFT, the SM should regularize at the high-energy cutoff $\Lambda_{\rm cut}$, fully preserving the SM gauge symmetries.  
A natural UV regularization is provided by a theory of new physics at $\Lambda_{\rm cut}$, for instance, quantum gravity. However, the theoretical inconsistency between the SM bilinear Lagrangian and the natural UV regularization, due to the No-Go theorem~\cite{NIELSEN1981173,NIELSEN1981219}, implies quadrilinear four-fermion operators and right-handed neutrinos, which effectively represent a theory of new physics at the UV cutoff. We adopt the four-fermion operators of the torsion-free Einstein-Cartan Lagrangian with SM fermion content and three right-handed neutrinos~\cite{Xue:2016dpl,Xue2017}:
\begin{equation}
    \mathcal{L}\supset-G_{\rm cut}\sum_{ff'}\big(\Bar{\psi}^{f}_{L}\psi^{f'}_{R}\Bar{\psi}^{f'}_{R}\psi^{f}_{L}+\Bar{\nu}^{fC}_{R}\psi^{f'}_{R}\Bar{\psi}^{f'}_{R}\nu^{fC}_{R}\big) + h.c. \label{EC4}
\end{equation}
where $\psi^{f}_{L}$ and $\psi^{f}_{R}$ (for simplicity $\psi^{f}_{R}$ is also used to represent the sterile neutrinos $\nu^{f}_{R}$) are the two-component Weyl fields that are gauge doublets and singlets of the symmetry $SU_{L}(2)\times U_{Y}(1)$, respectively. Fermion family indexes $f,f'=1,2,3$ are summed over for the three lepton families (charge $q=0,-1$) and for the three quark families (charge $q=2/3,-1/3$). Family mixing will be duly induced when one makes the unitary transformations $U_L$ and $U_R$ from gauge to mass eigenstates.
Attributing to the new physics at the cutoff, the effective four-fermion coupling $G_{\rm cut}\propto\mathcal{O}(\Lambda^{-2}_{\rm cut})$ is assumed to be unique for all the terms in Eq.~(\ref{EC4}), and it depends on the running energy 
scale $\mu$. Using strong-coupling $G_{\rm cut}\Lambda^2_{\rm cut}$ expansion to calculate two-point Green functions shows 
the presence of composite bosons $(\bar\psi_L\psi_R)$ and fermions $(\bar\psi_L\psi_R) \psi_L$ \cite{Xue1996}.     
An EFT for composite particles of masses $M\propto \Lambda$ is realized in the scaling domain of the stable UV fixed point at the composite scale $\Lambda$ ($v < \Lambda < \Lambda_{\rm cut}$) \cite{Xue2017}.
When the running energy scale $\mu$ decreases below $\Lambda$, the four-fermion interacting dynamics run into the SM ground state of spontaneous symmetry breaking and the IR fixed point of the $v$ \cite{Bardeen:1989ds}. In the IR scaling domain, integrating over massive composite states gives rise to effective four-fermion contact interactions of coupling $G\propto\mathcal{O}(\Lambda^{-2})$ \cite{Xue2022c}, for example, $G\big(\bar\psi^{i}_{La}t^a_{R}\big)\big(\bar t^{a}_{R}\psi_{Lia}\big)$ of the $\bar tt$-condensate model \cite{Bardeen:1989ds} in the third quark family. The one relevant for the physics case in exam is between quarks and leptons in the first SM family~\cite{Leonardi:2018jzn}
\begin{equation}
    \mathcal{L}_{q-\ell}\supset G\big(\Bar{l}^{i}_{L}N^{e}_{R}\big)\big(\Bar{u}^{a}_{R}\psi_{Lia}\big) + {h.c.} \label{contactt}
\end{equation}
Here $l^{i}_{L}=(\nu^{e}_{L},e_{L})$ and $e_{R}$ are lepton doublet and singlet of $SU(2)_{L}$, respectively, and $\psi_{Lia}=(u_{La},d_{La})$, $u_{Ra}$, $d_{Ra}$ are the counterparts for the quarks. The color $a$ and the weak isospin 
$i$ indexes are summed over.
$N^{e}_{R}$ represents the right-handed component of Majorana neutrino $N^e$ of mass $M_{N^e}$. On the basis of mass eigenstates, neglecting the mixing matrix, the contact interaction (\ref{contactt}) becomes
\begin{equation}
    \mathcal{L}_{q-\ell}\approx G\big(\Bar{e}P_{R}N^{e}\big)\big(\Bar{u}^{a}P_{L}d_{a}\big) + h.c. \quad , \label{contact}
\end{equation}
which is the one contributing to $n\to p+N+e$ decay.

On the other hand, the effective four-fermion contact interactions induce effective $W^\pm$ boson right-handed coupling \cite{Xue:2016dpl}
\begin{equation}
    \mathcal{L}\supset \mathcal{G}^{W}_{R}\Big(\frac{g_{w}}{\sqrt{2}}\Big)\big[(U^{\ell}_{R})^{\dagger}U^{\nu}_{R}\big]^{ll'}\Bar{l}\gamma^{\mu}P_{R}N^{l'}W^{-}_{\mu}+h.c.
    \label{gauge0}
\end{equation}
where ($N^{l}_{R},l_{R}$) are the 
right-handed leptons in the mass basis, $U^{\nu}_{R}$ and $U^{\ell}_{R}$ are unitary matrices $3\times 3$ ($\ell=e,\mu,\tau$). It's important to note that the mixing matrix 
$V_{\ell N}\equiv \big[(U^{\ell}_{R})^{\dagger}U^{\nu}_{R}\big]$ is not the PMNS one  $\big[(U^{\ell}_{L})^{\dagger}U^{\nu}_{L}\big]$. The effective operator of this type (\ref{gauge0}) contributes to vector boson fusion (VBF) processes, see the left Feynman diagram in Figure 1 of 
Ref.~\cite{CMS:2023nsv}.
The mixing $|V_{\ell N}|$ and mass $M_N$ are 
constrained \cite{CMS:2023nsv,Sirunyan2018,Sirunyan2019}. 
In this work, we neglect mixing by approximating $\big[(U^{\ell}_{R})^{\dagger}U^{\nu}_{R}\big]\approx 1$. Hence for the 
first lepton family, we have
\begin{equation}
    \mathcal{L}\approx \mathcal{G}^{W}_{R}\Big(\frac{g_{w}}{\sqrt{2}}\Big)\Bar{e}\gamma^{\mu}P_{R}N^{e}W^{-}_{\mu}+h.c. \label{gauge} \quad ,
\end{equation}
where $g_{w}$ is the normal $SU(2)_{L}$ coupling. The $\mathcal{G}^{W}_{R}$ parameterizes the effective $W^\pm$ right-handed coupling. The upper limit of its value should be smaller than $\sim 10^{-4}$. It is constrained by the $W^\pm$ decay width and regarding $N^e_R$ as a dark-matter particle \cite{Haghighat:2019rht, Shakeri:2020wvk}. 

The effective interactions (\ref{contact}) and (\ref{gauge}) are the relevant operators for computing the theoretical half-life of $0\nu\beta\beta$ decay, which is the aim of this article.

\section{Half-Lives Computations}
\label{sec:computations}

\begin{figure*}[ht!]\centering
\begin{subfigure}[t]{0.43\textwidth}            
\includegraphics[width=\linewidth]{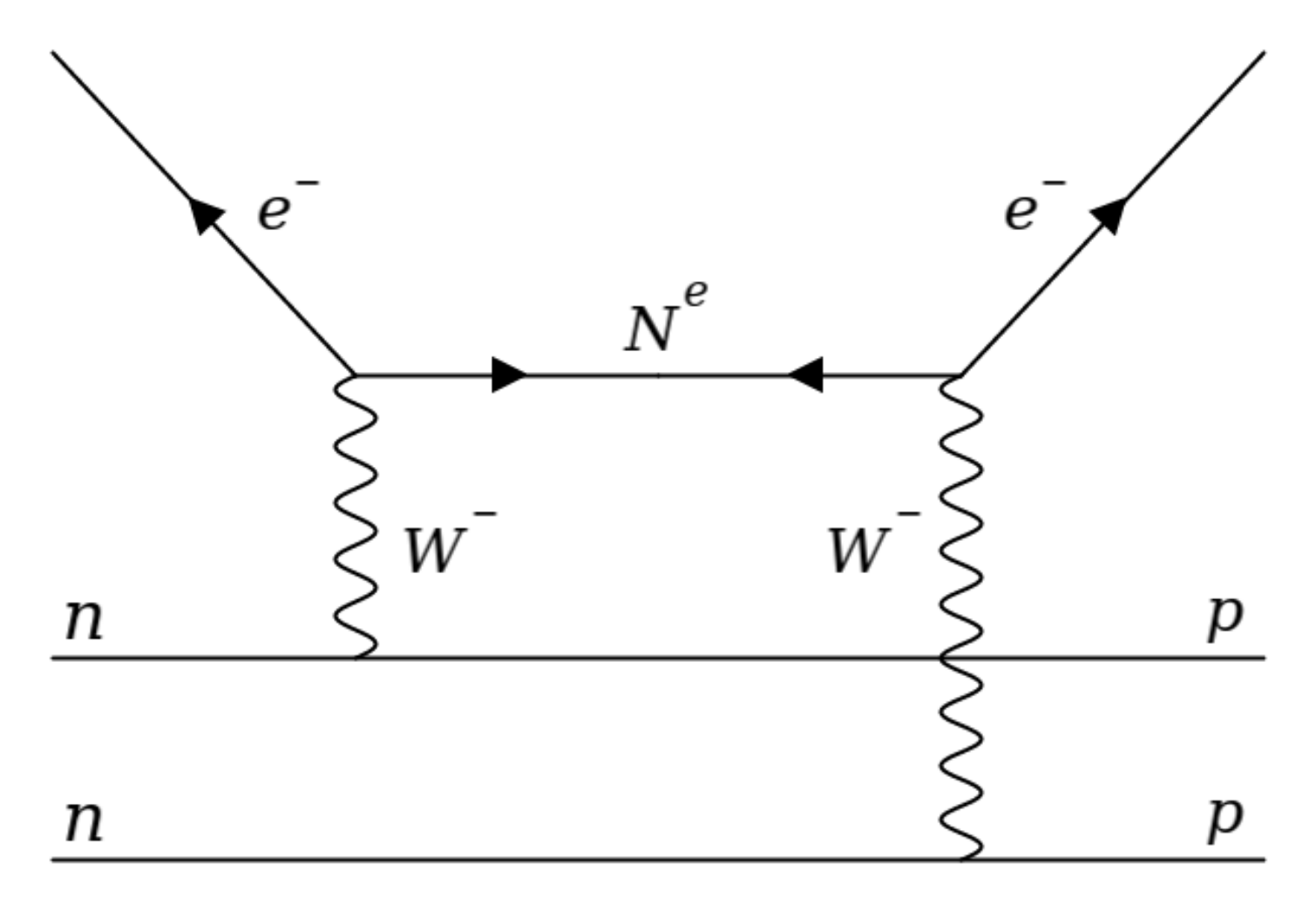}
\caption{Gauge term.}
\label{Fig:gauge}
\end{subfigure}
~~
\begin{subfigure}[t]{0.43\textwidth}
\includegraphics[width=\linewidth]{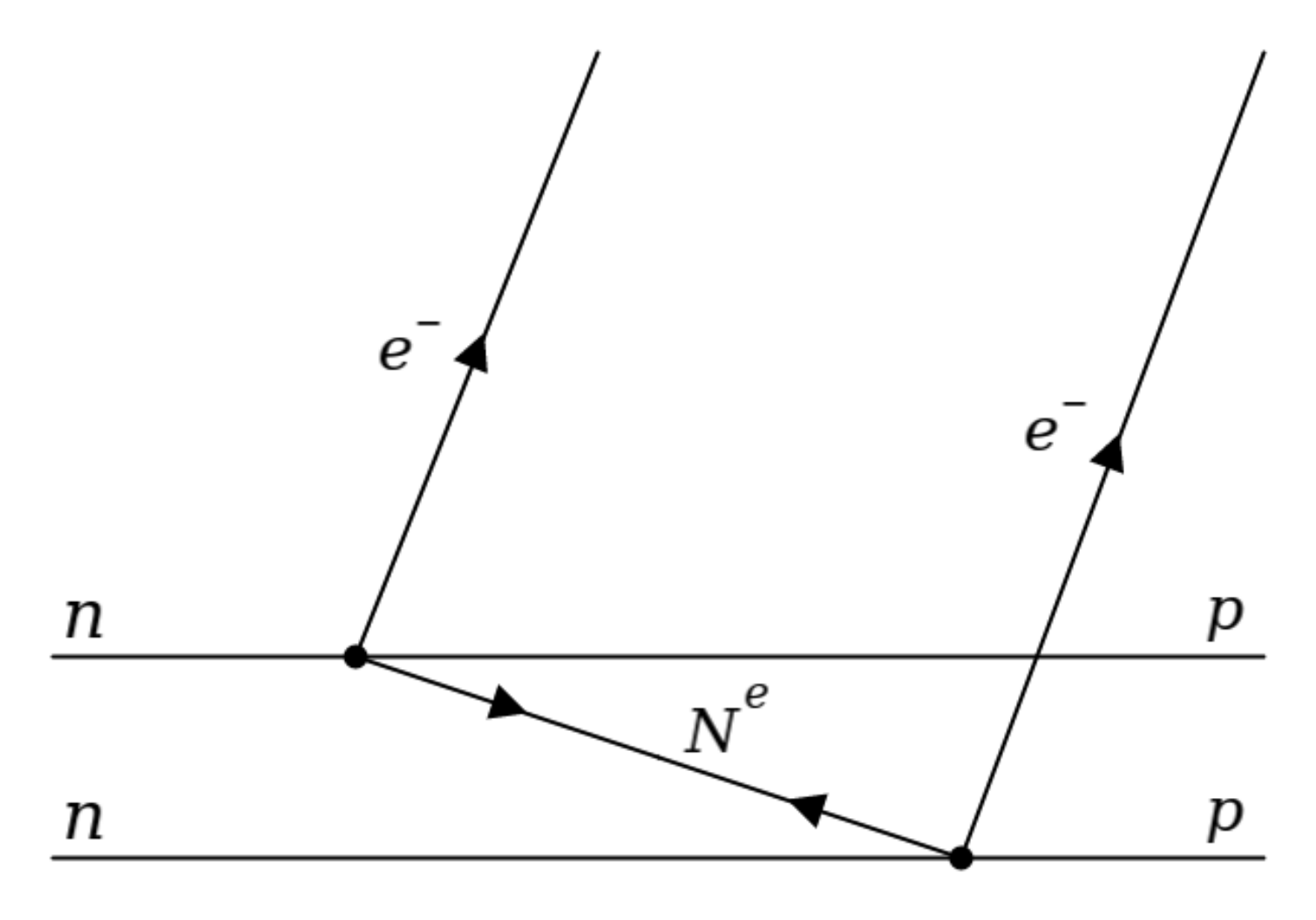}
\caption{Contact term.}
\label{Fig:contact}
\end{subfigure}\\
\begin{subfigure}[t]{0.43\textwidth}
\includegraphics[width=\linewidth]{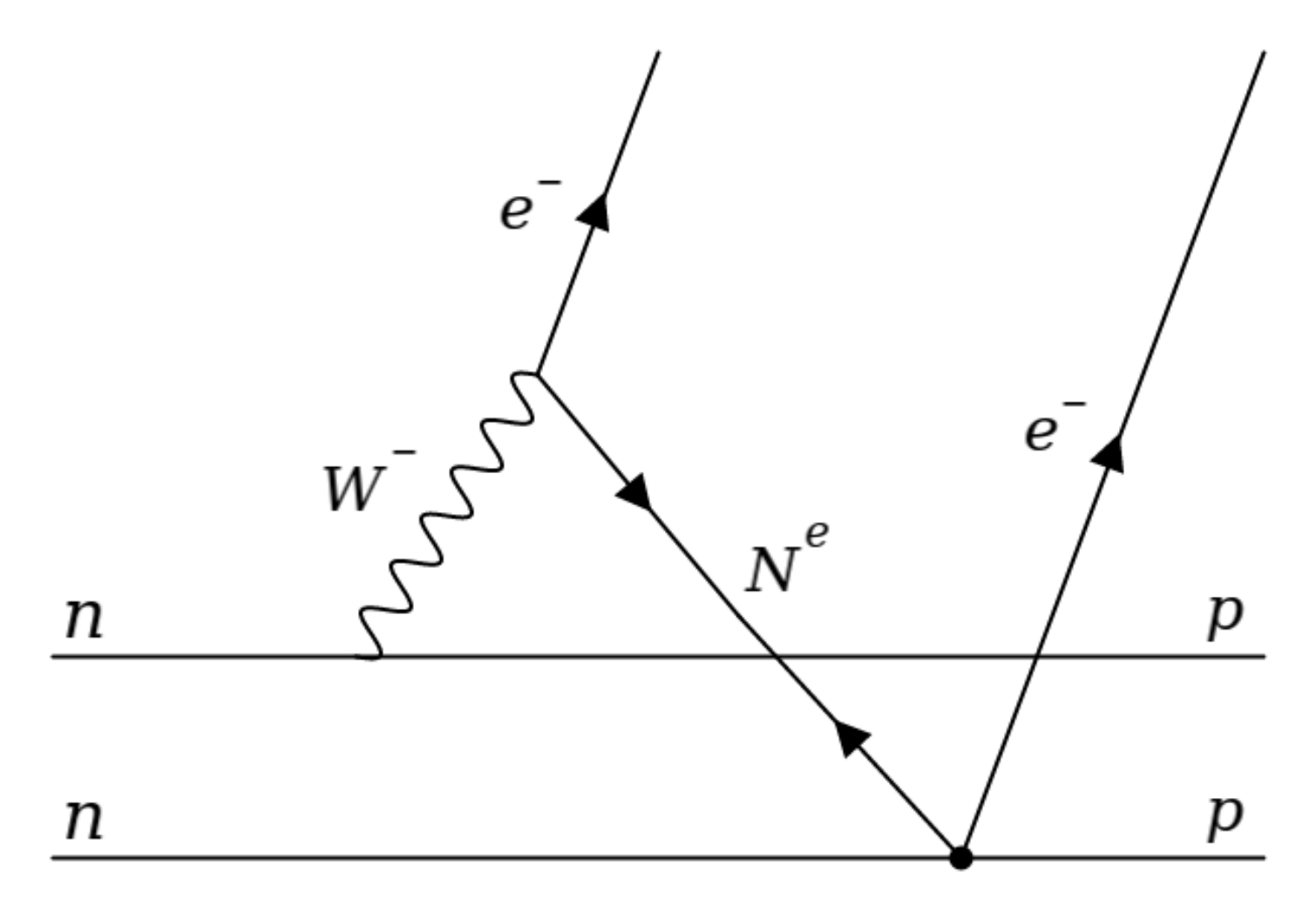}
\caption{Mixed term I.}
\label{Fig:mix1}
\end{subfigure}
~~
\begin{subfigure}[t]{0.43\textwidth}
\includegraphics[width=\linewidth]{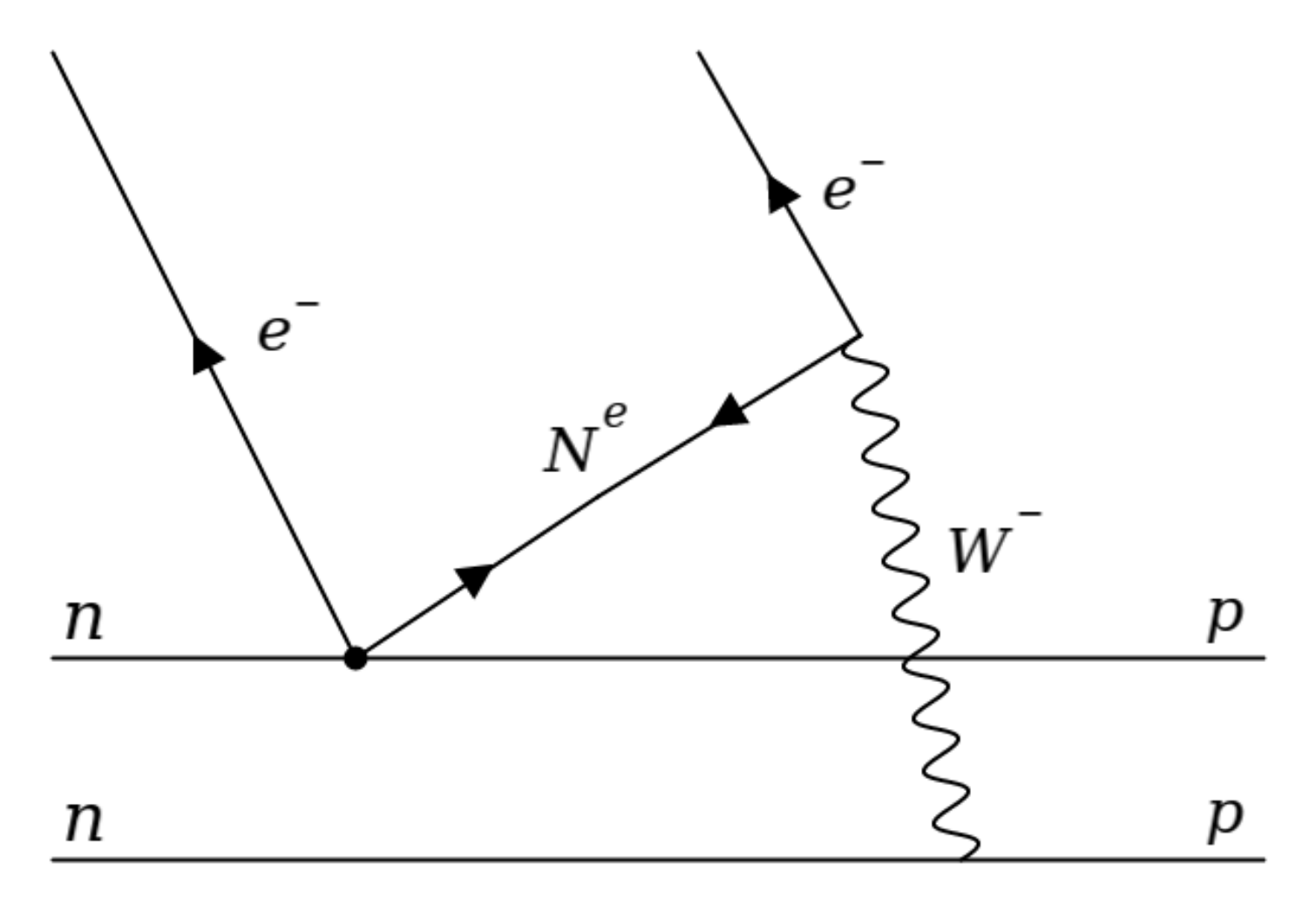}
\caption{Mixed term II.}
\label{Fig:mix2}
\end{subfigure}
\caption{Feynman diagrams for $0\nu\beta\beta$, i.e. the nuclear decay of two neutrons $n$ into two protons $p$, mediated by a Majorana neutrino. The two vertices connected by the exchanged heavy Majorana neutrino involve, from left to right, effective $V+A$ gauge interactions~(\ref{gauge}), four-fermion contact interactions~(\ref{contact}) and one gauge and one contact vertex, taking into account two permutations. }\label{Fig:diagrams}
\end{figure*}

In this section, we calculate the contributions of the effective contact interaction (\ref{contact}) and gauge interaction (\ref{gauge}) to the half-lives of the neutrinoless double beta decay. The relevant Feynman diagrams for $0\nu\beta\beta$, shown in Figure~\ref{Fig:diagrams}, with the exchange of a heavy Majorana neutrino, are four: one for pure gauge interactions~(Fig. \ref{Fig:gauge}), one for pure contact interactions~(Fig. \ref{Fig:contact}) and two mixed contributions~(Figs. \ref{Fig:mix1} and \ref{Fig:mix2}).

\subsection{Gauge Interactions}
In a pure gauge transition two neutrons $n$ decay in two protons $p$ emitting two $W^{-}$ bosons through a SM vertex:
\begin{equation}
    \frac{g_{w}}{2\sqrt{2}}\cos{\theta_{C}}J^{h}_{\mu}(x)W^{+\mu}(x)
\end{equation}
where $g_{w}$ is the $SU(2)_{L}$ SM gauge coupling, $\theta_{C}$ is the Cabibbo mixing angle ($\cos{\theta_{C}}\approx 0.974$), while $J^{h}_{\mu}(x)=\big(\Bar{u}\gamma_{\mu}(1-\gamma^{5})d\big)(x)$ is the hadronic weak charged current, where we have factored out $1/2$ from the chiral projector $P_{L}$ in order to conform with the expressions of nuclear matrix elements (NME) found in the literature~\cite{Barea:2009zza,Kotila:2021xgw,Simkovic:1999re,Deppisch:2020ztt}. The two $W$ bosons then decay into a right-handed electron, $e_{R}$, and a heavy Majorana electron neutrino, $N^{e}_{R}$, through the effective right interaction in Eq.~(\ref{gauge}). Summing up these two terms, we can write the gauge interactions Lagrangian as follows:
\begin{equation}
    \mathcal{L}_{\text{GI}}(x)=\frac{g_{w}}{\sqrt{2}}\Big[\mathcal{G}^{W}_{R}(j^{\mu}_{R}W^{-}_{\mu})(x)+\frac{1}{2}\cos{\theta_{C}}(J^{h}_{\mu}W^{+\mu})(x)\Big]+ h.c. \label{Lgi}
\end{equation}
with $P_{L,R}=\frac{1-\gamma^{5}}{2}$, while $j^{\mu}_{R}\equiv\Bar{e}\gamma^{\mu}P_{R}N^{e}$ is the leptonic current. 

For the purpose of this paper, we will consider only nuclear transitions between $0$-spin and positive parity states, i.e. $0^{+}\to 0^{+}$. The non-trivial \textit{S}-matrix element related to Eq.~(\ref{Lgi}) is 
\begin{align}
    T_{\text{GI}}=&-(\mathcal{G}^{W}_{R})^{2}\Big(\frac{g_{w}}{\sqrt{2}}\Big)^{4}\frac{(\cos{\theta_{C}})^{2}}{8}\Big(\frac{1-P_{12}}{\sqrt{2}}\Big)\int \frac{d^{4}q}{(2\pi)^{4}}d^{4}x d^{4}y e^{ip_{1}\cdot x}e^{ip_{2}\cdot y} \notag \\
    &\times e^{-iq\cdot (x-y)}\Bar{e}_{p_{1}}\gamma^{\mu}P_{R}\frac{i(\slashed{q}+M_{N})}{q^{2}-M^{2}_{N}+i\epsilon}P_{R}\gamma^{\nu}e^{C}_{p_{2}} \notag \\
    &\times \frac{(-i\delta^{\lambda}_{\mu})(-i\delta^{\rho}_{\nu})\braket{0^{+}_{F}|T\{J^{h}_{\lambda}(y)J^{h}_{\rho}(x)\}|0^{+}_{I}}}{\big((q+p_{1})^{2}-M^{2}_{W}+i\epsilon\big)\big((q-p_{2})^{2}-M^{2}_{W}+i\epsilon\big)} \quad , \label{contog3}
\end{align}
where $e_{p_{1,2}}$ are the wave functions of the electrons in the final state, with four-momenta $p_{1}$ and $p_{2}$, $q$ is the four-momentum transfer, i.e. the Majorana neutrino four-momentum, and $\frac{1-P_{12}}{\sqrt{2}}$ is the antisymmetric operator due to the production of two identical fermions. We make the \textit{ansatz} that the hadronic current is given by the sum of the nucleonic charged current 
\begin{equation}
    J^{h}_{\mu}(x)=\sum_{i}J^{(i)}_{\mu}(x) \label{ansatz}
\end{equation}
where the sum runs over the nucleons of the isotope which decays through $0\nu\beta\beta$. This implies that:
\begin{equation}
    \braket{0^{+}_{F}|T\{J^{h}_{\rho}(x)J^{h}_{\lambda}(y)\}|0^{+}_{I}}=e^{i(P_{F}-P_{I})\cdot y}\braket{0^{+}_{F}|T\{J^{h}_{\rho}(x-y)J^{h}_{\lambda}(0)\}|0^{+}_{I}}
\end{equation}
with $P_{I}$ and $P_{F}$ the four-momenta of the nucleus in the initial and final state, respectively. Now if we define $G_\text{eff}\equiv \mathcal{G}^W_{R}G_{F}\cos{\theta_{C}}$ and change the variables of integration as
\begin{equation}
    \begin{cases}
        x=z+\frac{u}{2} \\
         y=z-\frac{u}{2} 
    \end{cases} \quad \text{with} \quad d^{4}x d^{4}y=d^{4}u d^{4}z \label{subs}
\end{equation}
the integration over $d^{4}z$ gives the energy-momentum conservation, thus we can recast the matrix element in the form $T_{\text{GI}}\equiv i(2\pi)^{4}\delta^{4}\big(P_{F}+p_{1}+p_{2}-P_{I}\big)M_{\text{GI}}$, with
\begin{equation}
    M_{\text{GI}}=G^{2}_\text{eff}M^{4}_{W}M_{N}\frac{1-P_{12}}{\sqrt{2}}\int \frac{d^{4}q}{(2\pi)^{4}}\Bar{e}_{p_{1}}\gamma^{\mu}P_{R}\gamma^{\nu }e^{C}_{p_{2}}\frac{W_{\nu\mu}(q)}{\big(q^{2}_{0}-\omega^{2}_{N}+i\epsilon\big)\big(q^{2}_{0}-\omega^{2}_{W}+i\epsilon\big)^{2}} \label{MGIprov}
\end{equation}
in which we have neglected $p_{1}$ and $p_{2}$ with respect to $q$ and defined 
\begin{equation}
    W_{\nu\mu}(q)\equiv \int d^{4}u e^{-iu\cdot q}\braket{0^{+}_{F}|T\{J^{h}_{\nu}(u)J^{h}_{\mu}(0)\}|0^{+}_{I}} \quad , 
\end{equation}
where $\omega^{2}_{N(W)}=(\mathbf{q})^{2}+M^{2}_{N(W)}$. Using the identity: 
\begin{align}
    \frac{1-P_{12}}{\sqrt{2}}\Bar{e}_{p_{1}}\gamma^{\mu}P_{R}\gamma^{\nu }e^{C}_{p_{2}}=\frac{1}{\sqrt{2}}\big(\Bar{e}_{p_{1}}\gamma^{\mu}P_{R}\gamma^{\nu}e^{C}_{p_{2}}-\Bar{e}_{p_{2}}\gamma^{\mu}P_{R}\gamma^{\nu}e^{C}_{p_{1}}\big) \notag \\
    =\frac{1}{\sqrt{2}}\big(\Bar{e}_{p_{1}}\gamma^{\mu}\gamma^{\nu}P_{L}e^{C}_{p_{2}}+\Bar{e}_{p_{1}}\gamma^{\nu}\gamma^{\mu}P_{L}e^{C}_{p_{2}}\big)=\frac{2}{\sqrt{2}}g^{\mu\nu}\Bar{e}_{p_{1}}P_{L}e^{C}_{p_{2}} \quad ,
\end{align}
we obtain
\begin{equation}
    M_{\text{GI}}=\frac{2G^{2}_\text{eff}}{\sqrt{2}}M^{4}_{W}M_{N}\int \frac{d^{4}q}{(2\pi)^{4}}\Bar{e}_{p_{1}}P_{L}e^{C}_{p_{2}}\frac{W^{\mu}_{\mu}(q)}{\big(q^{2}_{0}-\omega^{2}_{N}+i\epsilon\big)\big(q^{2}_{0}-\omega^{2}_{W}+i\epsilon\big)^{2}} \quad . \label{MGIdef}
\end{equation}

As usual for $0\nu\beta\beta$ calculations~\cite{Panella:1997wa}, we insert a complete set of intermediate nuclear states $\ket{X}$ in $W^{\mu}_{\mu}$:
\begin{equation}
    W^{\mu}_{\mu}(q)=-i \int d^{3}\mathbf{u} e^{i\mathbf{u}\cdot\mathbf{q}}\sum_{\ket{X}}\Big(\frac{\braket{0^{+}_{F}|J^{h\mu}(\mathbf{u})|X}\braket{X|J^{h}_{\mu}(0)|0^{+}_{I}}}{q_{0}-E_{F}+E_{X}-i\epsilon} +\frac{\braket{0^{+}_{F}|J^{h}_{\mu}(0)|X}\braket{X|J^{h\mu}(\mathbf{u})|0^{+}_{I}}}{-q_{0}-E_{I}+E_{X}-i\epsilon}\Big) \ . \label{ob}
\end{equation}
The energy of a state $\ket{X}$ is $E_X=E_{CM}(\mathbf{P})+\epsilon_n$ where $E_{CM}(\mathbf{P})$ is the translational energy of the center of mass and $\epsilon_n$ is the excitation energy, characterizing state $\ket{X}$. Note also that the sum over intermediate states includes integration over the center of mass momentum $\mathbf{P}$ and a sum on the discrete part of the spectrum, i.e. the excitation level $n$. Now we use the closure approximation, namely we replace the intermediate state energies with an average value $\braket{E_X}=E_{CM}(\braket{\mathbf{P}})+\Bar{\epsilon}_{n}$, where $\Bar{\epsilon}_{n}$ is the average excitation energy of all the intermediate states $\ket{X}$. Due to the ansatz~(\ref{ansatz}), the center of mass motion ($\mathbf{R}=\frac{1}{A}\sum_{i}\mathbf{u}_{i}$) can be separated out so that we can rewrite the one body operator matrix elements~\cite{Lipkin:1955} in Eq.~(\ref{ob}) as 
\begin{equation}
    \braket{0^{+}_{F}|J^{h\mu}(\mathbf{u})|X}=\sum_{k}\braket{\braket{0^{+}_{F}|e^{i(\mathbf{P}-\mathbf{P}_{F})\cdot(\mathbf{u}-\mathbf{\xi}_{k})}\Tilde{J}^{(k)\mu}(\mathbf{P}-\mathbf{P}_{F})|X}} \label{hadron}
\end{equation}
and likewise for the other hadronic current. In~(\ref{hadron}) $\mathbf{\xi}_{i}=\mathbf{u}_{i}-\mathbf{R}$ are the relative coordinates, the notation $\braket{\bra{}\cdot\ket{}}$ denotes that we are in the space of the $A-1$ relative coordinates and $\Tilde{J}^{(k)\mu}(\mathbf{P}-\mathbf{P}_{F})$ is the nucleon current in momentum space. Hence, for the first term in~(\ref{ob}), we have:
\begin{equation}
    -i\int \frac{d^{3}\mathbf{P}}{(2\pi)^{3}}d^{3}\mathbf{u}e^{i\mathbf{q}\cdot\mathbf{u}}\sum_{k,\ell}\frac{e^{i(\mathbf{P}-\mathbf{P}_{F})\cdot(\mathbf{u}-\mathbf{\xi}_{k})}e^{i(\mathbf{P}-\mathbf{P}_{I})\cdot \mathbf{\xi}_{\ell}}}{q_{0}+\braket{E_{X}}-E_{F}-i\epsilon}\braket{\braket{0^{+}_{F}|\Tilde{J}^{(k)\mu}(\mathbf{P}-\mathbf{P}_{F})\Tilde{J}^{(\ell)}_{\mu}(-\mathbf{P}+\mathbf{P}_{I})|0^{+}_{I}}} \quad ,
\end{equation}
integrating over $d^{3}\mathbf{u}$ we obtain the Dirac delta $\delta^{(3)}(\mathbf{q}+\mathbf{P}-\mathbf{P}_{F})$, which can be used to integrate over the center of mass motion. Moreover, thanks to this Dirac delta and that of energy-momentum conservation we have the identities $\mathbf{P}-\mathbf{P}_{F}=-\mathbf{q}$ and $-\mathbf{P}+\mathbf{P}_{I}=\mathbf{q}$. We consider nuclei without recoil so that $E_{F}\approx M_{F}$, $E_{I}\approx M_{I}$ and $M_{F}\approx M_{I}$. By introducing the so-called closure energy~\cite{Tomoda:1990rs}
\begin{equation}
    \Delta=\braket{E_{X}}-\frac{1}{2}(M_{F}+M_{I})\approx 10 \text{ MeV} \quad ,
\end{equation}
we can recast the tensor function $W^{\mu}_{\mu}(q_0,\mathbf{q})$ in the form:
\begin{equation}
    W^{\mu}_{\mu}(q_{0},\mathbf{q})=\sum_{k,\ell}e^{i\mathbf{q}\cdot \mathbf{\xi}_{k\ell}}\braket{\braket{0^{+}_{F}|\Tilde{J}^{(k)\mu}(-\mathbf{q})\Tilde{J}^{(\ell)}_{\mu}(\mathbf{q})|0^{+}_{I}}}\frac{2i\Delta}{q^{2}_{0}-\Delta^{2}+i\epsilon} 
\end{equation}
where $\mathbf{\xi}_{k\ell}\equiv\mathbf{\xi}_{k}-\mathbf{\xi}_{\ell}$. Now we can perform the integration over $dq_0$ in Eq.~(\ref{MGIdef}) 
\begin{equation}
    I(q^{2}_{0})\equiv \int \frac{dq_{0}}{2\pi} \frac{2i\Delta}{(q^{2}_{0}-\Delta^{2})(q^{2}_{0}-\omega^{2}_{N})(q^{2}_{0}-\omega^{2}_{W})^{2}}=\frac{1}{\omega^{2}_{N}\omega^{4}_{W}}\approx \frac{1}{M^{2}_{N}M^{4}_{W}}
\end{equation}
where we have assumed $\Delta \ll\omega_W,\omega_N$ and that the exchanged Majorana neutrino is heavy, namely $M_N \gg 100 \text{ MeV}$ which is its momentum $|\mathbf{q}|\approx \frac{1}{r_{NN}}\approx 100 \text{ MeV}$ with $r_{NN}\approx 2 \text{ fm}$ the average inter-nucleon distance in the nuclei. 

Next, we use the non-relativistic limit for the four-vector nucleon currents~\cite{Kotila:2021xgw}:
\begin{equation}
    \Tilde{J}_{0}^{(k)}(\mathbf{q})=\tau^{+}_{k}\Big(F_{V}(\mathbf{q}^{2})\mathbb{1}_{2\times 2}+\frac{F_{P'}(\mathbf{q}^{2})}{4m^{2}_{p}}q_{0}(\mathbf{\sigma}_{k}\cdot \mathbf{q})+...\Big) \label{vecttemp}
\end{equation}
\begin{equation}
    \Tilde{J}^{(k)}_{i}(\mathbf{q})=\tau^{+}_{k}\Big(F_{A}(\mathbf{q}^{2})(\mathbf{\sigma}_{k})_{i}+\frac{F_{V}(\mathbf{q}^{2})+F_{W}(\mathbf{q}^{2})}{2m_{p}}i(\mathbf{\sigma}_{k}\times \mathbf{q})_{i}-\frac{F_{P'}(\mathbf{q}^{2})}{4m^{2}_{p}}q_{i}(\mathbf{\sigma}_{k}\cdot \mathbf{q})+...\Big) \label{vectvect}
\end{equation}
where $m_p$ is the proton mass, $\mathbf{\sigma}_{k}$ and $\tau^{+}_{k}$ are Pauli matrices acting respectively on the spin and isospin space of the $k$-th nucleon and $F_{o}$ are the nucleon form factors whose parametrizations and coupling constants $g_{o}$ are fixed as in Ref.~\cite{Kotila:2021xgw}. Lastly, taking into account only $0^+\to0^+$ transitions and keeping the first order in the non-relativistic limit we have 
\begin{align}
    M_{\text{GI}}&=\frac{2G^{2}_\text{eff}}{\sqrt{2}M_{N}}\Bar{e}_{p_{1}}P_{L}e^{C}_{p_{2}}\frac{m_{e}m_{p}}{4\pi R_{0}}g^{2}_{A}\Bigg[-\mathcal{M}_{GT}+\Big(\frac{g^{2}_{V}}{g^{2}_{A}}\Big)\mathcal{M}_{F}+\mathcal{M}_{T}\Bigg] \notag \\
    &=\frac{2G^{2}_\text{eff}}{\sqrt{2}M_{N}}\Bar{e}_{p_{1}}P_{L}e^{C}_{p_{2}}\frac{m_{e}m_{p}}{4\pi R_{0}}\mathcal{M}_{0\nu}
\end{align}
where $R_{0}=1.2A^{1/3} \text{ fm}$ is the mean nuclear radius, $m_e$ is the electron mass and $\mathcal{M}_{o}$ are the standard nuclear matrix elements (NME) for a $0\nu\beta\beta$ with heavy Majorana neutrino exchange~\cite{Simkovic:1999re,Kotila:2021xgw}.

The $0\nu\beta\beta$ half-life, which is the actual observable of the experimental searches~\cite{GERDA:2020xhi,KamLAND-Zen:2022tow}, formula is 
\begin{equation}
    (T^{0\nu}_{1/2})^{-1}_{\text{GI}}=\frac{1}{\ln{2}}\int \frac{d^{3}\mathbf{p}_{1}}{(2\pi)^{3}2E_{1}}\frac{d^{3}\mathbf{p}_{2}}{(2\pi)^{3}2E_{2}}\overline{|M_{\text{GI}}|^{2}}(2\pi)\delta(E_{F}+E_{1}+E_{2}-E_{I})
\end{equation}
where the probability amplitude squared and summed over the electron spin polarizations is
\begin{equation}
    \overline{|M_{\text{GI}}|^{2}}=\frac{G^{4}_\text{eff}}{8M^{2}_{N}}\frac{m^{2}_{e}m^{2}_{p}}{\pi^{2}R^{2}_{0}}|\mathcal{M}_{0\nu}|^{2}\sum_{spin}|\Bar{e}_{p_{1}}P_{L}e^{C}_{p_{2}}|^{2} \quad .
\end{equation}
We approximate the electron wave functions in the following way 
\begin{equation}
    e_{p_{1,2}}=\sqrt{F_{0}(Z+2,E_{1,2})} \ u(p_{1,2})
\end{equation}
where $F_{0}$ is the Fermi function, which describes the distortion of the electron wave function due to the Coulomb field of the nucleus~\cite{Doi:1982dn} and $u(p_{1,2})$ is the actual positive energy Dirac spinor. Now using simple Dirac algebra we obtain 
\begin{equation}
    \sum_{k,r}|\Bar{e}_{r,p_{1}}P_{L}e^{C}_{k,p_{2}}|^{2}=F_{0}(Z+2,E_{1})F_{0}(Z+2,E_{2})2p_{1}\cdot p_{2}
\end{equation}
so that, adopting standard notation~\cite{Deppisch:2020ztt} to express the phase-space integration 
\begin{equation}
    C=\frac{(G_{F}\cos{\theta_{C}})^{4}m^{2}_{e}}{16\pi^{5}}
\end{equation}
\begin{align}
    G^{(0)}_{11}&=\frac{2C}{(\ln{2})4R_{0}^{2}}\int F_{0}(Z+2,E_{1})F_{0}(Z+2,E_{2})p_{1}p_{2}E_{1}E_{2}\times \notag \\
    & \qquad \qquad \qquad \qquad \qquad \qquad \times dE_{1}dE_{2} \delta(E_{F}+E_{1}+E_{2}-E_{I}) \notag \\
    &\equiv \frac{2C}{(\ln{2})4R_{0}^{2}}\int f^{(0)}_{11+}p_{1}p_{2}E_{1}E_{2}dE_{1}dE_{2} \delta(E_{F}+E_{1}+E_{2}-E_{I})
\end{align}
we finally have the gauge interactions contributing to the $0\nu\beta\beta$ half-life 
\begin{equation}
    \big(T^{0\nu}_{1/2}\big)^{-1}_{\text{GI}}=(\mathcal{G}^{W}_{R})^{4}\frac{m^{2}_{p}}{M^{2}_{N}}|\mathcal{M}_{0\nu}|^{2}G^{(0)}_{11} \label{Tgauge}
\end{equation}

\subsection{Contact Interactions}
In this subsection, we go through the same type of computations we just did but for contact interactions, therefore for brevity's sake, we focus only on the biggest differences.

In a pure contact transition two neutrons $n$ decay simultaneously into two protons $p$, two electrons $e^{-}$ and exchange a heavy Majorana electron neutrino $N^{e}$ without mediating bosons
\begin{equation}
    \mathcal{L}_{\text{CI}}=G\big(\Bar{e}P_{R}N^{e}\big)\big(\Bar{u}^{a}P_{L}d_{a}\big)+h.c. \quad . \label{ci}
\end{equation}
From the NJL type model~\cite{Xue:2020cnw} we recall that the effective four-fermion coupling $G$ depends on the new physics energy scale and on the top-quark Yukawa coupling constant
\begin{equation}
    G\sim \frac{g^{2}_{t0}}{\Lambda^{2}}
\end{equation}
so that we can recast the Lagrangian in~(\ref{ci}) as follows:
\begin{align}
    \mathcal{L}_{\text{\text{CI}}}(x)&=\Big(\frac{g_{t0}}{\Lambda}\Big)^{2}\big(\Bar{e}P_{R}N^{e}\big)(x)\big(\Bar{u}^{a}P_{L}d_{a}\big)(x)+h.c. \notag \\
    &\equiv \Big(\frac{g^{2}_{t0}}{2\Lambda^{2}}\Big) j_{R}(x)J^{h}(x)+h.c. \quad , \label{Lci}
\end{align}
where in the last term we factored out the $1/2$ from the scalar hadronic current. 

The non-trivial $S$-matrix element related to Eq.~(\ref{Lci}) is
\begin{align}
    T_{\text{CI}}=&\Big(\frac{g^{4}_{t0}}{8\Lambda^{4}}\Big)\frac{1-P_{12}}{\sqrt{2}}\int \frac{d^{4}q}{(2\pi)^{4}}d^{4}x d^{4}y e^{ip_{1}\cdot x}e^{ip_{2}\cdot y}e^{-iq\cdot (x-y)} \notag \\
    &\times \Bar{e}_{p_{1}}P_{R}\frac{i(\slashed{q}+M_{N})}{q^{2}-M^{2}_{N}+i\epsilon}P_{R}e^{C}_{p_{2}}\braket{0^{+}_{F}|T\{J^{h}(x)J^{h}(y)\}|0^{+}_{I}} \quad , \label{Smatc}
\end{align}
where the main difference with the pure gauge case is that the leptonic and hadronic currents are now scalar. Making the same substitutions as in~(\ref{subs}), the \textit{ansatz}~(\ref{ansatz}), simplifying the leptonic currents with simple Dirac algebra and defining the scalar function 
\begin{equation}
    W(q)\equiv \int d^{4}u e^{-iq\cdot u}\braket{0^{+}_{F}|T\{J^{h}(u)J^{h}(0)\}|0^{+}_{I}} 
\end{equation}
we find:
\begin{equation}
    M_{\text{CI}}=\Big(\frac{g^{4}_{t0}}{4\sqrt{2}\Lambda^{4}}\Big)M_{N}\int \frac{d^{4}q}{(2\pi)^{4}}\Bar{e}_{p_{1}}P_{R}e^{C}_{p_{2}}\frac{W(q)}{q^{2}_{0}-\omega^{2}_{N}+i\epsilon} \quad . \label{MCIdef}
\end{equation}
Here too we will compute the function $W(q)$ inserting a complete set of intermediate nuclear states $\ket{X}$ and adopting the closure approximation in order to obtain the NMEs. Hence we find:
\begin{equation}
    W(q_{0},\mathbf{q})=\sum_{k,\ell}e^{i\mathbf{q}\cdot \mathbf{\xi}_{k\ell}}\braket{\braket{0^{+}_{F}|\Tilde{J}^{(k)}(-\mathbf{q})\Tilde{J}^{(\ell)}(\mathbf{q})|0^{+}_{I}}}\frac{2i\Delta}{q^{2}_{0}-\Delta^{2}+i\epsilon} \quad .
\end{equation}
We perform the integration over $dq_0$ in Eq.~(\ref{MCIdef})
\begin{equation}
    \int \frac{dq_{0}}{2\pi} \frac{2i\Delta}{(q^{2}_{0}-\Delta^{2})(q^{2}_{0}-\omega_{N}^{2})}=\frac{1}{\omega_{N}(\Delta+\omega_{N})}\approx \frac{1}{M^{2}_{N}}
\end{equation}
where we have assumed $\Delta\ll \omega_N$ and that the exchanged Majorana neutrino is heavy, i.e. $M_N\gg 100\text{ MeV}$. 

Next, we use the non-relativistic limit for the scalar nucleon currents~\cite{Kotila:2021xgw}, which is the main difference from the pure gauge case:
\begin{equation}
    \Tilde{J}^{(k)}(\mathbf{q})=\tau^{+}_{k}\Big(F_{S}(\mathbf{q}^{2})\mathbb{1}_{2\times2}-\frac{F_{P}(\mathbf{q}^{2})}{2m_{p}}(\mathbf{\sigma}_{k}\cdot \mathbf{q})+ ... \Big) \quad . \label{hadrscal}
\end{equation}
Taking into account only $0^+\to0^+$ transitions and keeping the first order in the non-relativistic limit we have 
\begin{align}
    M_{\text{CI}}&=\Big(\frac{g^{4}_{t0}}{4\sqrt{2}\Lambda^{4}}\Big)\frac{1}{M_{N}}\Bar{e}_{p_{1}}P_{R}e^{C}_{p_{2}}\frac{m_{e}m_{p}}{4\pi R_{0}}\Big[g^{2}_{V}\mathcal{M}_{F}-\frac{g^{2}_{P'}}{12}\big(\mathcal{M}'^{PP}_{GT}+\mathcal{M}'^{PP}_{T}\big)\Big] \notag \\
    &\equiv\Big(\frac{g^{4}_{t0}}{4\sqrt{2}\Lambda^{4}}\Big)\frac{1}{M_{N}}\Bar{e}_{p_{1}}P_{R}e^{C}_{p_{2}}\frac{m_{e}m_{p}}{4\pi R_{0}}\mathcal{M}_{1} \quad .
\end{align}
The $0\nu\beta\beta$ half-life formula is 
\begin{equation}
    \big(T^{0\nu}_{1/2}\big)^{-1}_{\text{CI}}=\frac{1}{\ln{2}}\int \frac{d^{3}\mathbf{p}_{1}}{(2\pi)^{3}2E_{1}}\frac{d^{3}\mathbf{p}_{2}}{(2\pi)^{3}2E_{2}} \overline{|M_{\text{CI}}|^{2}}(2\pi)\delta(E_{F}+E_{1}+E_{2}-E_{I})
\end{equation}
where the amplitude squared and summed over the electron spin polarizations is 
\begin{equation}
    \overline{|M_{\text{CI}}|^{2}}=\Big(\frac{g_{t0}}{\Lambda}\Big)^{8}\frac{m^{2}_{e}m^{2}_{p}}{512\pi^{2} R^{2}_{0}M^{2}_{N}}|\mathcal{M}_{1}|^{2}\sum_{spin}|\Bar{e}_{p_{1}}P_{R}e^{C}_{p_{2}}|^{2} \quad .
\end{equation}
We approximate the electron wave functions with the Fermi function~\cite{Doi:1982dn} and, using simple Dirac algebra, we obtain
\begin{equation}
    \sum_{k,r}|\Bar{e}_{r,p_{1}}P_{R}e^{C}_{k,p_{2}}|^{2}=F_{0}(Z+2,E_{1})F_{0}(Z+2,E_{2})2p_{1}\cdot p_{2} 
\end{equation}
so that, adopting standard notation~\cite{Deppisch:2020ztt} to express the phase-space integration 
\begin{equation}
    C=\frac{(G_{F}\cos{\theta_{C}})^{4}m^{2}_{e}}{16\pi^{5}}
\end{equation}
\begin{align}
    G^{(0)}_{11}&=\frac{2C}{(\ln{2})4R_{0}^{2}}\int F_{0}(Z+2,E_{1})F_{0}(Z+2,E_{2})p_{1}p_{2}E_{1}E_{2}\times \notag \\
    & \qquad \qquad \qquad \qquad \qquad \qquad \times dE_{1}dE_{2} \delta(E_{F}+E_{1}+E_{2}-E_{I}) \notag \\
    &\equiv \frac{2C}{(\ln{2})4R_{0}^{2}}\int f^{(0)}_{11+}p_{1}p_{2}E_{1}E_{2}dE_{1}dE_{2} \delta(E_{F}+E_{1}+E_{2}-E_{I})
\end{align}
we finally have the contact interactions contribution to the $0\nu\beta\beta$ half-life
\begin{equation}
    \big(T^{0\nu}_{1/2}\big)^{-1}_{\text{CI}}=\Big(\frac{g^{8}_{t0}}{\Lambda^{8}}\Big)\frac{m^{2}_{p}}{64M^{2}_{N}}|\mathcal{M}_{1}|^{2}\frac{G^{(0)}_{11}}{(G_{F}\cos{\theta_{C}})^{4}} \label{Tcontact} \quad .
\end{equation}

\subsection{Mixed Interactions}
To conclude this section, it is mandatory to consider the case where one neutron decays through gauge interaction~(\ref{Lgi}) and the other through contact interaction~(\ref{Lci}) and vice versa: the so-called mixed diagrams (Figs.~\ref{Fig:mix1} and~\ref{Fig:mix2}). Usually, the mixed interactions' contributions to $0\nu\beta\beta$ half-life cancel each other out or, in general, are negligible compared to the pure cases~\cite{Biondini:2021vip}. However, with this NJL-type model, we will show that there is constructive interference between the two mixed terms. 

For brevity's sake, the computations and considerations are only expressed for $T^{I}_{\text{mix}}$ (Fig.~\ref{Fig:mix1}) since those for $T^{II}_{\text{mix}}$ (Fig.~\ref{Fig:mix2}) are perfectly analogous. The non-trivial $S$-matrix element related to Fig.~\ref{Fig:mix1} is 
\begin{align}
    T^{I}_{\text{mix}}=&C_\text{eff}\int \frac{d^{4}q}{(2\pi)^{4}}d^{4}xd^{4}y e^{-iq\cdot (x-y)}e^{ip_{1}\cdot x}e^{ip_{2}\cdot y}\frac{1-P_{12}}{\sqrt{2}} \notag \\
    &\times \Bar{e}_{p_{1}}\gamma^{\mu}P_{R}\frac{i(\slashed{q}+M_{N})}{q^{2}-M^{2}_{N}+i\epsilon}P_{R}e^{C}_{p_{2}}\frac{(-i\delta)^{\nu}_{\mu}\braket{0^{+}_{F}|T\{J^{h}(x)J^{h}_{\nu}(y)\}|0^{+}_{I}}}{(q+p_{1})^{2}-M^{2}_{W}+i\epsilon}
\end{align}
where $C_\text{eff}\equiv\mathcal{G}^{W}_{R}\cos{\theta_{C}}(\frac{g_{t0}}{\Lambda})^{2}\frac{M^{2}_{W}G_{F}}{2\sqrt{2}}$. Again, with the same assumptions, substitutions and ansatz of the previous subsections, we obtain:
\begin{equation}
    M^{I}_{\text{mix}}=-iC_\text{eff}M_{N}\int \frac{d^{4}q}{(2\pi)^{4}}\frac{1-P_{12}}{\sqrt{2}}\frac{\Bar{e}_{p_{1}}\gamma^{\mu}P_{R}e^{C}_{p_{2}}}{q^{2}_{0}-\omega^{2}_{N}+i\epsilon}\frac{W_{\mu}(q-p_{1})}{q_{0}^{2}-\omega^{2}_{W}+i\epsilon} \quad . \label{MI}
\end{equation}
The action of the antisymmetric operator over the leptonic states is as follows:
\begin{equation}
    \frac{1-P_{12}}{\sqrt{2}}\Bar{e}_{p_{1}}\gamma^{\mu}P_{R}e^{C}_{p_{2}}=\frac{1}{\sqrt{2}}\big(\Bar{e}_{p_{1}}\gamma^{\mu}P_{R}e^{C}_{p_{2}}-\Bar{e}_{p_{1}}\gamma^{\mu}P_{L}e^{C}_{p_{2}}\big) \quad . \label{mixI}
\end{equation}
Instead for the other $S$-matrix element related to Fig.~\ref{Fig:mix2} we can obtain
\begin{equation}
    M^{II}_{\text{mix}}=iC_\text{eff}M_{N}\int \frac{d^{4}q}{(2\pi)^{4}}\frac{1-P_{12}}{\sqrt{2}}\frac{\Bar{e}_{p_{1}}P_{R}\gamma^{\mu}e^{C}_{p_{2}}}{q^{2}_{0}-\omega^{2}_{N}+i\epsilon}\frac{W_{\mu}(q-p_{1})}{q_{0}^{2}-\omega^{2}_{W}+i\epsilon} \quad , \label{MII}
\end{equation}
where the action of the antisymmetric operator gives 
\begin{equation}
    \frac{1-P_{12}}{\sqrt{2}}\Bar{e}_{p_{1}}P_{R}\gamma^{\mu}e^{C}_{p_{2}}=\frac{1}{\sqrt{2}}\big(\Bar{e}_{p_{1}}P_{R}\gamma^{\mu}e^{C}_{p_{2}}-\Bar{e}_{p_{1}}P_{L}\gamma^{\mu}e^{C}_{p_{2}}\big)
\end{equation}
which is identical to the action of the other antisymmetric operator. 

Now if we add together the two contributions~(\ref{MI}) and~(\ref{MII}) we have:
\begin{equation}
    M_{\text{mix}}\equiv iC_\text{eff}\frac{M_{N}}{\sqrt{2}}\int \frac{d^{4}q}{(2\pi)^{4}}-2\Bar{e}_{p_{1}}\gamma^{\mu}\gamma^{5}e^{C}_{p_{2}}\frac{W_{\mu}(q)}{(q^{2}_{0}-\omega^{2}_{N}+i\epsilon)(q_{0}^{2}-\omega^{2}_{W}+i\epsilon)}
\end{equation}
so we can't neglect a priori the significance of the mixed diagrams. Let us briefly show how NMEs are obtained in this case; by applying the closure approximation we can recast the four-vector function $W_{\mu}(q)$ as follows 
\begin{equation}
    W_{\mu}(q_{0},\mathbf{q})=\sum_{k,\ell}e^{i\mathbf{q}\cdot \mathbf{\xi}_{k\ell}}\braket{\bra{0^{+}_{F}}|\Tilde{J}^{(k)}(-\mathbf{q})\Tilde{J}^{(\ell)}_{\mu}(\mathbf{q})|0^{+}_{I}}\frac{2i\Delta}{q^{2}_{0}-\Delta^{2}+i\epsilon} \label{Wmu}
\end{equation}
and performing the integration in $q_0$ we obtain
\begin{equation}
    \int \frac{dq_{0}}{2\pi}\frac{2i\Delta}{(q^{2}_{0}-\Delta^{2})(q^{2}_{0}-\omega^{2}_{N})(q^{2}_{0}-\omega^{2}_{W})}=-\frac{1}{\omega^{2}_{W}\omega^{2}_{N}}\approx -\frac{1}{M^{2}_{W}M^{2}_{N}} \quad ,
\end{equation}
so that we have
\begin{equation}
    M_{\text{mix}}=i\mathcal{G}^{W}_{R}\Big(\frac{g^{2}_{t0}}{\Lambda^{2}}\Big)\frac{G_{F}\cos{\theta_{C}}}{2M_{N}}\int \frac{d^{3}\mathbf{q}}{(2\pi)^{3}}\Bar{e}_{p_{1}}\gamma^{\mu}\gamma^{5}e^{C}_{p_{2}}\sum_{k,\ell}e^{i\mathbf{q}\cdot \mathbf{\xi}_{k\ell}}\braket{\bra{0^{+}_{F}}|\Tilde{J}^{(k)}(-\mathbf{q})\Tilde{J}^{(\ell)}_{\mu}(\mathbf{q})|0^{+}_{I}} \ . \label{Mmix}
\end{equation}
Then we use the already seen non-relativistic expansions for the scalar (\ref{hadrscal}) and four-vector ((\ref{vecttemp}) and (\ref{vectvect})) nucleon currents~\cite{Kotila:2021xgw} in order to perform the integration in $\mathbf{q}$:
\begin{align}
    M_{\text{mix}}&=i\mathcal{G}^{W}_{R}\Big(\frac{g^{2}_{t0}}{\Lambda^{2}}\Big)\frac{G_{F}\cos{\theta_{C}}}{2M_{N}}\Bar{e}_{p_{1}}\gamma^{0}\gamma^{5}e^{C}_{p_{2}}\frac{m_{e}m_{p}}{4\pi R_{0}} \notag \\
    & \qquad \times \Big[g_{S}g_{V}\mathcal{M}_{F}+\frac{g_{P}g_{P'}}{24}\big(\mathcal{M}'^{q_{0}PP'}_{GT}+\mathcal{M}'^{q_{0}PP'}_{T}\big)\Big] \notag \\
    &\equiv i\mathcal{G}^{W}_{R}\Big(\frac{g^{2}_{t0}}{\Lambda^{2}}\Big)\frac{G_{F}\cos{\theta_{C}}}{2M_{N}}\Bar{e}_{p_{1}}\gamma^{0}\gamma^{5}e^{C}_{p_{2}}\frac{m_{e}m_{p}}{4\pi R_{0}}\mathcal{M}_{5} \label{Mmixdef}
\end{align}
where, from reference~\cite{Deppisch:2020ztt}, $\mathcal{M}'^{q_{0}PP'}_{GT}\approx 10^{-2}\mathcal{M}'^{P'P'}_{GT}$ and $\mathcal{M}'^{q_{0}PP'}_{T}\approx 10^{-2}\mathcal{M}'^{P'P'}_{T}$, which would allow mixed contributions to be neglected. However the coupling constants $g_{P}$ and $g_{P'}$ are of the order of $10^2$ whereas the others are all of the order of $1$, so mixed diagrams give a contribution of the same order as gauge and contact ones. It is important to point out in Eq.~(\ref{Mmixdef}) that only the $\gamma^0$ matrix survived integration in $\mathbf{q}$ because it is the only one that contributes to the mix of scalar and four-vector nuclear currents to the first perturbative order. 

The $0\nu\beta\beta$ half-life formula is 
\begin{equation}
    \big(T^{0\nu}_{1/2}\big)^{-1}_{\text{mix}}=\frac{1}{\ln{2}}\int \frac{d^{3}\mathbf{p}_{1}}{(2\pi)^{3}2E_{1}}\frac{d^{3}\mathbf{p}_{2}}{(2\pi)^{3}2E_{2}}\overline{|M_{\text{mix}}|^{2}}(2\pi)\delta(E_{F}+E_{1}+E_{2}-E_{I}) \label{emivita_mix}
\end{equation}
where the amplitude squared and summed over the electron spin polarizations is 
\begin{equation}
    \overline{|M_{\text{mix}}|^{2}}=\big(\mathcal{G}^{W}_{R}\big)^{2}\Big(\frac{g_{t0}}{\Lambda}\Big)^{4}\frac{(G_{F}\cos{\theta_{C}})^{2}}{4M^{2}_{N}}\frac{m^{2}_{e}m^{2}_{p}}{(4\pi R_{0})^{2}}|\mathcal{M}_{5}|^{2}\sum_{spin}|\Bar{e}_{p_{1}}\gamma^{0}\gamma^{5}e^{C}_{p_{2}}|^{2} \quad .
\end{equation}
Lastly, we approximate the electron wave functions with the Fermi function~\cite{Doi:1982dn} and with simple Dirac algebra we obtain
\begin{equation}
    \sum_{k,r}|\Bar{e}_{p_{1}}\gamma^{0}\gamma^{5}e^{C}_{p_{2}}|^{2}=F_{0}(Z+2,E_{1})F_{0}(Z+2,E_{2})4(E_{1}E_{2}+m^{2}_{e}+\mathbf{p}_{1}\cdot \mathbf{p}_{2}) 
\end{equation}
so that, with standard notation~\cite{Deppisch:2020ztt} to express the phase-space integration
\begin{equation}
    C=\frac{(G_{F}\cos{\theta_{C}})^{4}m^{2}_{e}}{16\pi^{5}}
\end{equation}
\begin{align}
    G^{(0)}_{66}&=\frac{2C}{(\ln{2})4R_{0}^{2}}\int F_{0}(Z+2,E_{1})F_{0}(Z+2,E_{2})\frac{(2E_{1}E_{2}+2m^{2}_{e})}{4E_{1}E_{2}} \notag \\
    & \qquad \qquad \qquad \qquad \quad \times p_{1}p_{2}E_{1}E_{2}dE_{1}dE_{2} \delta(E_{F}+E_{1}+E_{2}-E_{I}) \notag \\
    &\equiv \frac{2C}{(\ln{2})4R_{0}^{2}}\int \frac{1}{16}f^{(0)}_{66}p_{1}p_{2}E_{1}E_{2}dE_{1}dE_{2} \delta(E_{F}+E_{1}+E_{2}-E_{I})
\end{align}
we have the non-zero mixed interactions contribution to the $0\nu\beta\beta$ half-life
\begin{equation}
    \big(T^{0\nu}_{1/2}\big)^{-1}_{\text{mix}}=\big(\mathcal{G}^{W}_{R}\big)^{2}\Big(\frac{g^{4}_{t0}}{\Lambda^{4}}\Big)\frac{m^{2}_{p}}{2M^{2}_{N}}|\mathcal{M}_{5}|^{2}\frac{G^{(0)}_{66}}{(G_{F}\cos{\theta_{C}})^{2}} \label{Tmix} \quad .
\end{equation}

\section{Bounds on the Model Parameters}
\label{sec:results}

Finally, here we will put together the results of the last section to obtain the constraints on the BSM model's parameters. Assuming that the three contributions~(\ref{Tgauge}),~(\ref{Tcontact}) and~(\ref{Tmix}) are dominant to $0\nu\beta\beta$ half-life, we have
\begin{equation}
    T^{0\nu}_{1/2}\equiv (T^{0\nu}_{1/2})_{\text{GI}}+(T^{0\nu}_{1/2})_{\text{CI}}+(T^{0\nu}_{1/2})_{\text{mix}} \quad , \label{Tipotesi}
\end{equation}
and by imposing the experimental lower bounds on $0\nu\beta\beta$, due to the non-observation of this rare decay, Eq.~(\ref{Tipotesi}) becomes the following inequality:
\begin{align}
    (T^{0\nu}_{1/2})_{\text{exp}}\leq& \ \frac{M^2_{N}}{(\mathcal{G}^{W}_{R})^{4}}\Bigg(\frac{1}{m^2_{p}|\mathcal{M}_{0\nu}|^2G^{(0)}_{11}}\Bigg)+\frac{\Lambda^4M^2_{N}}{(\mathcal{G}^W_R)^2} \notag \\
    &\times \Bigg(\frac{2(G_F\cos{\theta_C})^2}{g^4_{t0}m^2_p|\mathcal{M}_{5}|^2G^{(0)}_{66}}\Bigg)+\Lambda^8M^2_{N}\Bigg(\frac{64(G_F\cos{\theta_C})^4}{g^8_{t0}m^2_p|\mathcal{M}_{1}|^2G^{(0)}_{11}}\Bigg) \label{inequality}
\end{align}
which is the main analytical result of the paper. To obtain the explicit constraint we insert the appropriate values for the NMEs ($\mathcal{M}_{0\nu}$, $\mathcal{M}_{1}$, $\mathcal{M}_{5}$ from Tables V and VI of reference~\cite{Deppisch:2020ztt}, with IBM model), the Phase Space Factors, PSFs, ($G^{(0)}_{11}$, $G^{(0)}_{66}$ from Table VII of reference~\cite{Deppisch:2020ztt}) and for the other quantities. For the top-quark Yukawa coupling constant $g_{t0}\sim\mathcal{O}(1)$ as suggested in reference~\cite{Xue:2020cnw}. 

Up to date, the most stringent experimental bounds on $0\nu\beta\beta$ are:
\begin{equation}
    (T^{0\nu}_{1/2})_{\text{exp}}=
    \begin{cases}
        1.8\times 10^{26} \text{ yr} \quad (^{76}\text{Ge~\cite{GERDA:2020xhi})} \\
        2.3\times 10^{26} \text{ yr} \quad (^{136}\text{Xe~\cite{KamLAND-Zen:2022tow}})
    \end{cases} \quad ,
\end{equation}

\begin{figure*}[ht]
\centering
\begin{subfigure}[]{0.48\textwidth}            
\includegraphics[width=\textwidth]{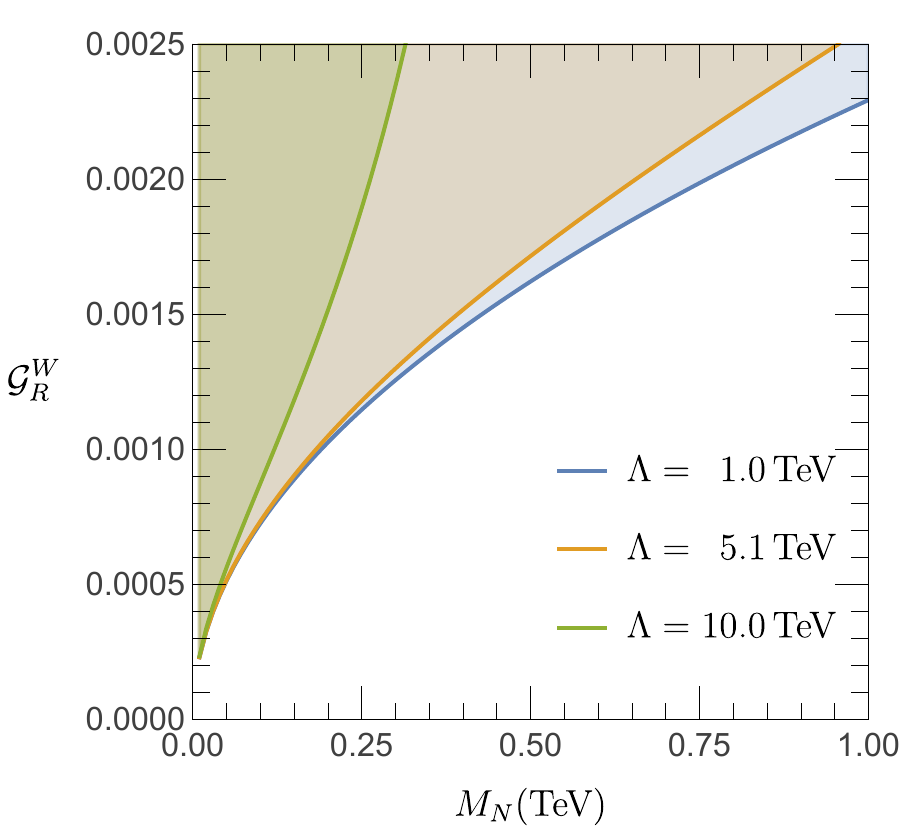}
\caption{}
\label{Fig:bounds1}
\end{subfigure}
~
\begin{subfigure}[]{0.48\textwidth}
\includegraphics[width=\textwidth]{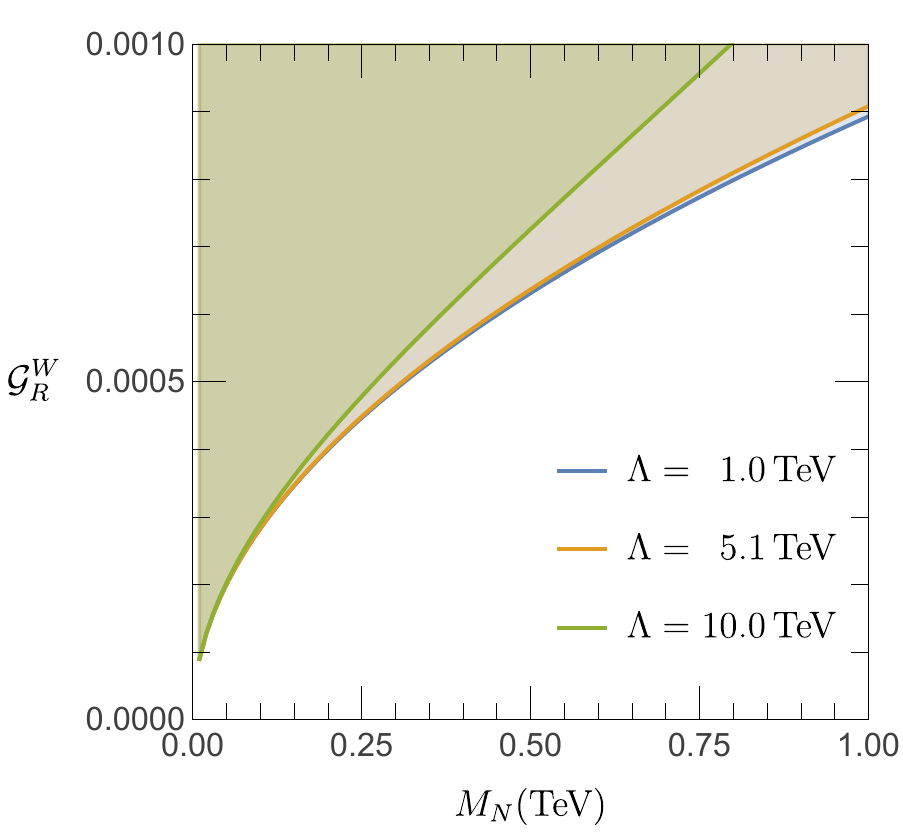}
\caption{}
\label{Fig:bounds2}
\end{subfigure}
\caption{Bounds on $\mathcal{G}^W_R$ as a function of the heavy Majorana neutrino mass $M_N$, for fixed values of $\Lambda$. The regions excluded by Eq.~(\ref{inequality}) are those colored above the three curves. Note that the plots start from the reference value $M_N=10 \text{ GeV ($0.01$ TeV)}$, because of the heavy Majorana neutrino approximation. (a) Constraints obtained with $0\nu\beta\beta$ lower bound $(T^{0\nu}_{1/2})=2.3\times 10^{26}\text{ yr}$ from KamLAND-Zen experiment. (b) Constraints obtained with expected $0\nu\beta\beta$ lower bound $(T^{0\nu}_{1/2})\sim 10^{28}\text{ yr}$ from next generation experiments.}\label{Fig:bounds}
\end{figure*}
so that the analysis of~(\ref{inequality}) will be done with $^{76}$Ge and $^{136}$Xe isotopes. Nevertheless, the constraints obtained with $^{76}$Ge will be less stringent than with $^{136}$Xe, so we ignore the former. 

By assigning reference values to the energy scale $\Lambda$ ($\Lambda=1$ TeV, $\Lambda=5.1$ TeV as suggested in~\cite{Xue:2020cnw}, $\Lambda=10$ TeV), we can extract from~(\ref{inequality}) an upper bound for the effective coupling $\mathcal{G}^{W}_{R}$ as a function of the heavy Majorana neutrino mass $M_{N}$, as shown in Fig.~\ref{Fig:bounds1} with colored areas representing the excluded regions. The constraint imposed by the non-observation of the $0\nu\beta\beta$ decay when $\Lambda=10 \text{ TeV}$ (green curve) becomes very weak before reaching $M_N=1 \text{ TeV}$, while for the other values of $\Lambda=1, 5.1 \text{ TeV}$ (blue and orange curves, respectively) the limits obtained for the effective coupling constant are $\mathcal{G}^{W}_{R}\lesssim 2.6\times 10^{-3}$ and $\mathcal{G}^{W}_{R}\lesssim 2.3\times 10^{-3}$ at $M_N=1 \text{ TeV}$.

In a similar way, it is possible to foresee the bounds on the effective right coupling $\mathcal{G}^W_R$ coming from future $0\nu\beta\beta$ experiments~\cite{LEGEND:2021bnm,nEXO:2021ujk,Nakamura:2020szx}. The next generation of $0\nu\beta\beta$ experiments aims at sensitivities for the half-life of the order of $(T^{0\nu}_{1/2})_{\text{exp}}\sim 10^{28} \text{ yr} \ (^{136}\text{Xe projection~\cite{nEXO:2021ujk}})$. 
The future exclusion limits are given in Fig.~\ref{Fig:bounds2}. The bounds are much stronger and will constrain $\mathcal{G}^W_R$ below $10^{-3}$ for the whole parameter space considered, with an improvement of a factor $\sim 2.5$ at the benchmark point $M_N = \Lambda = 1 \text{ TeV}$.

The results presented in this manuscript are valid only for Majorana neutrino masses heavier than the typical momentum transfer for $0\nu\beta\beta$ reactions, i.e. $M_N\gg 100 \text{ MeV}$. Therefore, limit curves in Fig.~\ref{Fig:bounds} are presented for the reference value of $M_N=10 \text{ GeV} \gg 100 \text{ MeV}$.
We do not address the light neutrinos regime ($M_N < 100 \text{ MeV}$) since the computation would rely on different approximations, such as the use of NMEs dependent on the mass of the neutrinos. 
We focus on a mass range up to $M_N=1$ TeV since it implies small effective right coupling constant $\mathcal{G}^{W}_{R}$ values, which are preferred in this model~\cite{Xue:2020cnw}. In addition, the chosen mass range is particularly appealing as it is often associated with high-energy searches for heavy neutrinos at colliders.

Finally, the same effective right coupling $\mathcal{G}^W_R$ has been studied 
in reference~\cite{Shakeri:2020wvk} from stellar cooling data~\cite{Diaz:2019kim}. We find that our constraints are, in general, stronger. 

\section{Summary and Remarks}
\label{sec:conclusion}

In this paper, we studied the constraints on the parameters of a BSM model of NJL type arising from the non-observation of the rare $0\nu\beta\beta$ decay. We presented the NJL type EFT~\cite{Xue:2016dpl,Leonardi:2018jzn} with its distinctive, quark-quark and quark-lepton, four-fermion operators which we have shown can contribute to the half-life of $0\nu\beta\beta$~\cite{derMateosian:1964vza,Fiorini:1967in,Dolinski:2019nrj}. After that, we computed these gauge, contact, and mixed contributions by using the ansatz and approximations typical of double beta decay calculations~\cite{Doi:1982dn}. Hence we exploited the non-observation of $0\nu\beta\beta$ from state-of-the-art experiments and translated the newly calculated theoretical half-life into a constraint in the 3D parameter space ($M_{N}$, $\Lambda$, $\mathcal{G}^{W}_{R}$) of the examined BSM model. 

By setting some values, of the order of magnitude suggested by the reference~\cite{Xue:2020cnw}, for the New Physics energy scale $\Lambda$, we converted the 3D constraint into an upper bound, dependent on the mass of the heavy Majorana neutrino $M_{N}$, for the effective coupling constant $\mathcal{G}^{W}_{R}$. This right coupling constant is that of vertexes of the SM gauge boson $W$ and right-handed currents, which means that in our physical scenario, they are non-vanishing and parity symmetry could be restored at the energy scale $\Lambda\sim \mathcal{O}(\text{TeV})$. Furthermore, we used the expected half-life limits from the next-generation experiments to foresee the future projected upper bound on $0\nu\beta\beta$. 
Finally, we confirmed the consistency of our constraints through observed stellar cooling data. 

If we include the flavor mixing $V_{\ell N}\equiv \big[(U^{\ell}_{R})^{\dagger}U^{\nu}_{R}\big]$ (\ref{contactt}) between three SM families, the final results will receive the contributions of Majorana neutrinos $N^\mu$ ($M_{N^\mu}$) and $N^\tau$ ($M_{N^\tau}$) from the second and third families. Namely, effective right-handed interacting vertexes $\sim \mathcal{G}^{W}_{R} V_{e N^{\mu,\tau}} W_\mu\bar e \gamma^\mu P_R N^{\mu,\tau}$ are present in Figs. \ref{Fig:gauge}, \ref{Fig:mix1} and \ref{Fig:mix2}. In these cases, the effective coupling becomes $(\mathcal{G}^{W}_{R})^2|V_{e N^{\mu,\tau}}|^2<(\mathcal{G}^{W}_{R})^2$ because of $|V_{e N^{\mu,\tau}}|^2<1$. We do not know the mixing $V_{\ell N}$, masses $M_{N^e}$, $M_{N^\mu}$ and $M_{N^\tau}$ of sterile Majorana neutrinos $N^e$, $N^\mu$ and $N^\tau$. In this study of $0\nu\beta\beta$ process, Figures \ref{Fig:bounds1} and \ref{Fig:bounds2} show constraints on the general parameter space of the effective right-handed coupling $\mathcal{G}^{W}_{R}|V_{e N^{e,\mu,\tau}}|$ and sterile Majorana neutrinos' masses $M_{N^{e,\mu,\tau}}$. The results represent actually contributions from all sterile Majorana neutrinos $N^e$, $N^\mu$, and $N^\tau$, because of nontrivial mixing $V_{\ell N}\not=0$. $N^e$ could be light and a candidate for warm DM particles in the present Universe. While massive $N^\mu$ and $N^\tau$ could have already decayed to SM particles in the early Universe. However, they could dominantly contribute to the $0\nu\beta\beta$ process via Feynman diagrams of Figure \ref{Fig:diagrams}. To disentangle contributions from sterile Majorana neutrinos of three SM families requires necessarily different experiments and observations for different lepton final states.
Therefore, it is worthwhile to study the phenomenology of the effective operators (\ref{contact}) and (\ref{contactt}), relating to the experiments of Probing Heavy Majorana Neutrinos and the Weinberg Operator through Vector Boson Fusion Processes in Proton-Proton Collisions \cite{CMS:2023nsv}, the experiment of MiniBooNE Neutrino Oscillation \cite{MiniBooNE:2018esg, MiniBooNE:2020pnu}, and the measurement of the Positive Muon Anomalous Magnetic Moment \cite{Muong-2:2021ojo}.



\acknowledgments

The work of M.~P. was supported by the Alexander von Humboldt-Stiftung.




\bibliography{references}{}
\bibliographystyle{JHEP.bst}

\end{document}